# Plasmon mode excitation and photoluminescence enhancement on silver nanoring

A.A. Kuchmizhak[1], S.O. Gurbatov,[1] Yu.N. Kulchin,[1,2] O.B. Vitrik[1,2]

[1]Institute of Automation and Control Processes, Far Eastern Branch, Russian Academy of Science, Vladivostok 690041, Russia
[2]Far Eastern Federal University, 8 Sukhanova str., Vladivostok 690041, Russia
*Corresponding author: ku4mijak@dvo.ru

We demonstrate a simple and high-performance laser-assisted technique for silver nanoring fabrication, which includes the ablation of the Ag film by focused nanosecond pulses and subsequent reactive ion polishing. The nanoring diameter and thickness can be controlled by optimizing both the pulse energy and the metal film thickness at laser ablation step, while the subsequent reactive ion polishing provides the ability to fabricate the nanoring with desirable height. Scattering patterns of s-polarized collimated laser beam obliquely illuminating the nanoring demonstrate the focal spot inside the nanoring shifted from its center at a distance of ~ 0.57$R_{ring}$. Five-fold enhancement of the photoluminescence signal from the Rhodamine 6G organic dye near the Ag nanoring was demonstrated. This enhancement was attributed to the increase of the electromagnetic field amplitude near the nanoring surface arising from excitation of the multipole plasmon modes traveling along the nanoring. This assumption was confirmed by dark-field back-scattering spectrum of the nanoring measured under white-light illumination, as well as by supporting finite-difference time-domain simulations.

## 1. Introduction

Noble metal ring nanostructures are increasingly attracting great interest because they provide a significant enhancement of localized electromagnetic field due to surface plasmons excitation, and also allow for fine-tuning of plasmon resonances through variation of ring size and shape [1-4]. Such nanostructures have already found their way into many areas: surface enhanced spectroscopy (SERS, SEIRA) as highly effective antennas [5], optical data storage as recording marks to achieve super resolution [6], the optical telecommunications band as plasmonic waveguides [7] as well as in biological and chemical sensors [8]. To produce ring metal nanostructures, methods of colloidal [2], electron [3], and ion beam lithography [5] as well as chemical synthesis [4] are applied, which enables us to create single nanorings of various size, geometry, and form, as well as their periodic arrays. However, technologically much simpler laser surface modification techniques can be found having proved the ability to create a wide range of functional nanostructures [9-17].

This paper will discuss that using such methods in conjunction with subsequent reactive ion etching, which will allow us to produce individual ring Ag-nanostructures with controlled geometrical dimensions on a dielectric substrate. Also it will be shown that such nanorings can increase the electromagnetic field intensity near their surface similar to the structures made by other techniques. Application of the laser-produced nanorings as functional plasmonic nanostructure will be demonstrated for local enhancement of the photoluminescence signal.

## 2. Nanoring fabrication

Fabrication of submicron rings is revealed to be done in two stages. At the first stage the 35- and 50-nm-thick silver films deposited by e-beam evaporation (Ferrotec EV M-6) on the smooth quartz substrate are normally irradiated by single second-harmonic (532 nm) pulses (pulse duration ~ 7 ns, maximum pulse energy – 10 mJ) of a Nd:YAG laser system. Each pulse was focused onto the Ag film surface using the focusing (NA=0.65) objective (see details in Fig. 1 and in Ref. [14]). To improve the quality of the laser beam used for nanostructuring each pulse was coupled to the section of the single-mode optical fiber (Thorlabs SM400), which acts as a spatial filter and provides almost perfect Gaussian-like energy distribution on its output. The pulse energy was modulated by a variable filter and controlled using high sensitive photodetector (J-10SI-HE Energy Sensor, Coherent EPM2000).

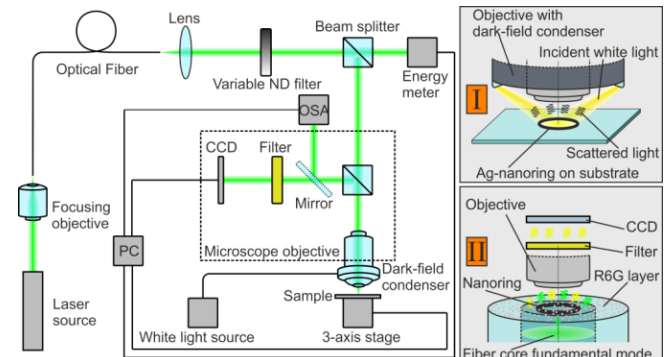

Fig. 1. Schematic of the experimental setup used for laser nanostructuring of Ag films and studying the optical properties of the structures obtained. Inset I schematically shows the dark-field illumination and scattering signal detection from the Ag nanoring. Inset II shows the optical scheme of the experiment on the photoluminescence enhancement of organic dye Rhodamine 6G.

Laser pulse impact on the surface of metal films with thicknesses $h_{film}$=35 nm and $h_{film}$=50 nm at variable pulse energies E results in formation of the ring-shaped structures representing through holes surrounded by a resolidified melt rim (Figs. 2a and 2c, respectively). It should be noted that such structures appear only under ns-pulse irradiation, while femtosecond pulse irradiation does not provide resolidified melt rings surrounded the through holes (in this case femtosecond Ti:Sapphire-laser system is used as a radiation source in Fig. 1). Apparently, in the latter case Ag film detaches from the substrate due to thermoelastic tensions caused by the temperature-driven gradients [15,18], which results in formation of the microhole surrounded by the hemispherical cupola-like walls (Fig. 2e). On the contrary, nanosecond pulses initiate other thermodynamic processes caused by subsurface boiling in the film-substrate interface, which results in spreading of the melted metal and its solidification in the form of the ring with a smooth uniform edge at the through hole border [19,20]. Figures 2a and 2c also show that additional crown-like nanospikes at the nanoring edge appear under ns-pulse irradiation (E > 41 nJ for $h_{film}$=35 nm and E > 52 nJ for $h_{film}$=50 nm), which is an undesired effect in terms of achieving the regular geometrical shape of the nanoring.

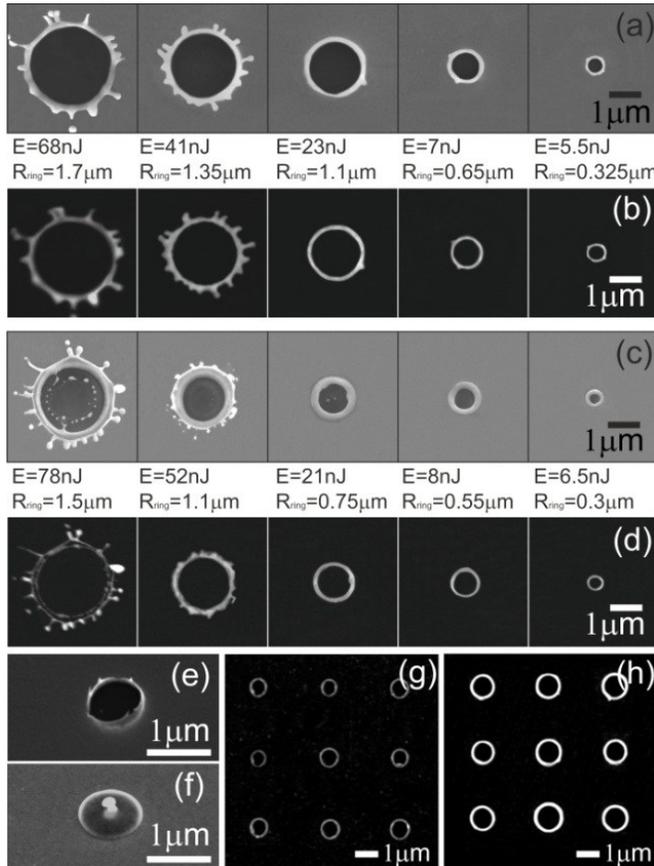

Fig.2 SEM images of laser-induced through holes in 35- and 50-nm-thick Ag films obtained at variable pulse energies E before (a,c) and after polishing with accelerated Ar+ ions (b,d); (e)-(f) Side-view (at 45°) SEM images of the through microhole fabricated in the 50-nm-thick Ag film under single fs-pulse irradiation and the structure resembling the nanojet fabricated in the 50-nm-thick Ag film under single ns-pulse irradiation at pulse energy E=4,1 nJ; (g,h)- SEM images of the square regular arrays of the Ag nanoring with the radii of 350nm and 550nm.

However, at lower pulse energies E (E < 37 nJ for $h_{film}$=35 nm and E < 45 nJ for $h_{film}$=50 nm) the nanorings demonstrate regular, nearly annular, shape with the radii of the structures $R_{ring}$ ranging from 0.33 to 1.10 μm for the 35-nm-thick film and from 0.30 to 0.75 μm for 50-nm-thick film. As seen, the minimal diameter of the fabricated nanorings at such film thicknesses determining the minimum possible distance between adjacent nanoelements arranged into the array can achieve 0.3 μm. At pulse energies lower than 5 nJ for 35-nm-thick film and 6 nJ for 35-nm-thick film the through microholes and, as a consequence, the nanoring do not form. Instead, structures representing well-known resolidified nanojets (Fig. 2f) appear under ns-pulse irradiation [9]. In accordance with atomic force microscope, the height of all regular-shaped rings exceeds 200 nm, which is significantly higher than the initial Ag film thickness providing the ability to fabricate separately lying nanorings on the dielectric substrate via layer-by-layer polishing of the surrounded Ag film. Reactive ion polishing (RIP, Hitachi I4000) providing an average rate ~0.33 nm/sec is used for this purpose. In order to minimize the possible melting of the metal film under the action of the heating Ar+ beam, the RIP procedure was performed for several consecutive polishing cycles each of which does not exceed 15 seconds followed by metal film cooling during 1 min after each cycle.

Figures 2 b,d demonstrate the result of thinning the nanostructures previously presented in Figures 2 a,c for the total ion beam exposure times 110 and 155 sec, respectively. As seen, such polishing removes the metal film in the vicinity of the nanorings without any significant distortions of the regular ring shape. The height h of the obtained rings can be varied by optimizing the RIP exposure time. In this work the height of the nanorings was chosen to be 200 nm and 250 nm for structures fabricated in the 35- and 50-nm-thick Ag films, respectively. Likewise, the ring thickness t was found to be increased with initial film thickness. The resulting nanoring thicknesses were 150 nm and 200 nm for 35- and 50-nm-thick films, respectively. To demonstrate the reproducibility of the developed fabrication technique, square regular arrays of nanorings with $R_{ring}$=350nm, t=80nm, h=100nm (Fig. 2g) and $R_{ring}$=550nm, t=200nm, h=200nm (Fig. 2h), respectively, were fabricated on the 50-nm-thick Ag film. The observed variation of ring diameter $\Delta R_{ring}$ in these arrays can be explained by the pulse energy fluctuation at the 5-% level and does not exceed 70 nm and 100 nm for 35- and 50-nm-thick films, respectively. Apparently, $\Delta R_{ring}$ value can be reduced by improving the laser source stability.

## 3. Optical properties of the nanoring

The metal nanoring with the outer radius $R_{ring}$ significantly greater than the ring thickness t ($R_{ring}$>>t) represents an antenna-like nanostructure amplifying incident electromagnetic radiation via the strong coupling to the circumferentially traveling surface plasmon modes [5,21]. Such nanostructure also can be used to efficiently convert the incoming propagating radiation into the strong evanescent waves [22]. However, along with the abovementioned advantages laser-based technique discussed in this paper almost excludes the possibility to fabricate metal nanorings with a uniform cross-section along their length (see Fig. 2). In this paper, we demonstrate that this feature does not significantly affect the ability of the nanorings to concentrate and enhance the electromagnetic fields. For this purpose, we fabricate the silver nanoring with the outer radius $R_{ring}$ = 1.1 µm, mean thickness t=150 nm and the height h=100 nm (See Fig. 3a and Ref. [23] for details) directly on the endface of a 10-mm-long section of the single-mode optical fiber (OF, Thorlabs S460-HP, [24]). Accurate positioning of the laser beam used to ablate the 35-nm-thick silver film on the OF endface provides the ability to fabricate the nanoring with the abovementioned geometrical dimensions within the area of its 3.8-µm-wide core (Fig. 3a). This solution significantly simplifies the mechanical manipulation with the nanoring as well as provides an additional opportunity of its excitation by almost ideal Gaussian beam, which was inputted to the OF from the opposite endface (at laser wavelengths >470 nm, [24]). We shall show that this feature has proved useful for the photoluminescent image detection.

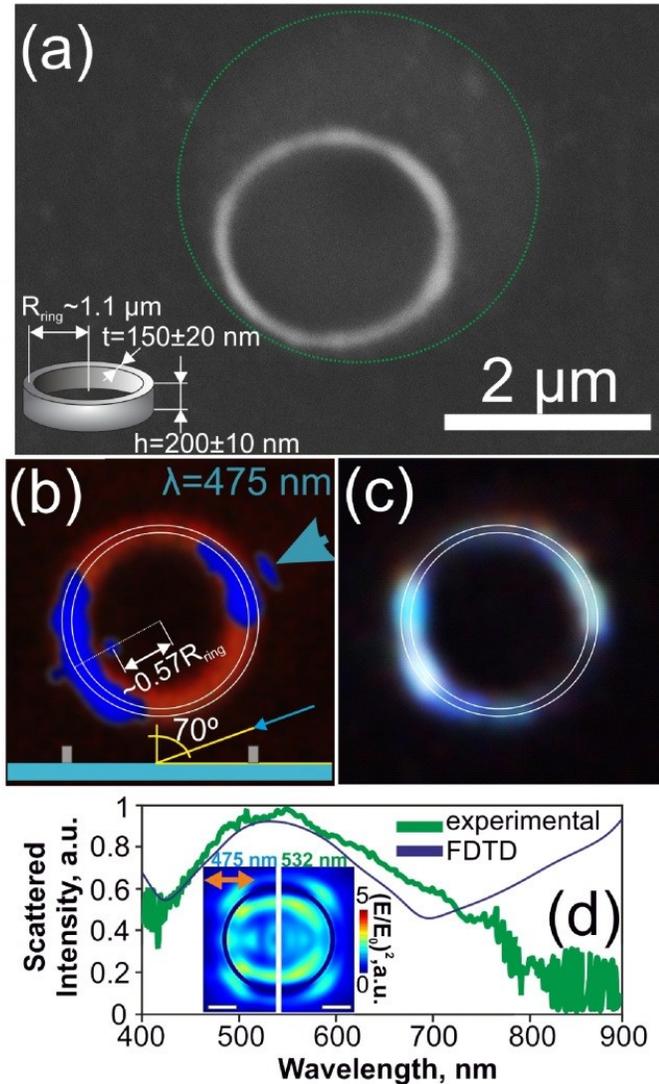

Fig.3. Scattering properties of silver nanoring. (a) Side-view (at an angle of 45°) SEM image of the silver nanoring fabricated on the endface of the optical fiber within the fiber core area (Green circle indicates the actual fiber core size. Inset schematically shows the real geometrical dimensions of the nanoring). (b) Optical image demonstrating the characteristic scattering of the s-polarized laser radiation (λ=475 nm) on the Ag nanoring obtained at oblique irradiation (20°) of the nanoring (Blue arrow indicates the irradiation direction). (c) Dark-field back-scattering image of the nanoring irradiated by unpolarized white light source (White circles show the actual nanoring size). (d) Measured and calculated dark-field scattering spectra of the Ag nanoring. Insets show the squared normalized electric-field amplitude $(E/E_0)^2$ near the nanoring calculated at λ=475 nm (left) and 532 nm (right). (Polarization direction is marked by the double-direction orange arrow. Scale bars correspond to 600 nm).

Characteristic scattering pattern of s-polarized collimated beam from the semiconductor continuous wave (CW) laser with a center wavelength λ = 475 nm obliquely illuminating the nanoring at an angle of 70° with respect to the fiber endface normal (Fig. 3b) clearly demonstrates the focal spot inside the nanoring shifted from its center at a distance of ~ 0.6$R_{ring}$. It should be noted that this result is in good qualitatively and quantitatively agreement with the results obtained for nanorings with significantly larger radii (up to 10 microns, [21,25]), demonstrating the universal character of the oblique laser radiation scattering on such nanostructures. Note that similar scattering patterns are observed under oblique irradiation of the nanoring using different wavelengths (not shown in this paper) demonstrating that the fabricated nanorings despite some geometric shape imperfections exhibit optical properties similar to the properties of both nanorings fabricated by e-beam lithography and chemically synthesized single-crystalline Ag nanorings with almost perfect geometric shape [5,21,22,25].

Presence of such focal spot indicates the constructive interference of laser radiation scattered from the nanoring surface. However, excitation of the high-order evanescent modes [22] as well as dark plasmons [5] possibly can contribute to the scattering pattern. However, owing to sufficiently large nanoring size the contribution of such near-field components to the scattering pattern appears to be negligible. This conclusion is partly confirmed by the fact that almost identical scattering patterns are observed on both metal and dielectric nanorings [3,22] of the same size, although near-field contributions for these rings to the scattering signal and focal spot formation obviously must be different.

Dark-field back-scattering (see. Inset in Fig. 1a for details) image of the nanoring obtained under white-light illumination indicates the excitation of plasmon modes in the visible spectral range, which appears as bright blue-green scattered light (Fig. 3c). It should be noted that due to the high aspect ratio $R_{ring}/t$ of the Ag nanoring, the fundamental dipole resonance is strongly shifted in the IR spectral region and thus cannot contribute to the visible scattering.

To verify our assumptions concerning the spectral composition of the scattered radiation we measured the dark-field scattering spectrum of the nanoring. Scattered radiation from the nanoring illuminated by the white light source was collected by the optical microscope objective (NA = 0.95, 100x) and directed onto the spectrometer (Andor Shamrock 303i) equipped with the sensitivity TE-cooled EMCCD-camera (Newton 971, [26]). Scattered spectrum of the nanoring normalized on both the spectrum measured from the flat surface of the OF endface and the white light source spectrum is shown in Fig. 3d. As can be seen, the experimentally measured scattering demonstrates the broad peak (in the range 480-580 nm), which is consistent with the observed dark field scattering (Fig. 3c).

To provide a better understanding of the experimental results, we carried out the supporting numerical simulation of the optical radiation scattering on the nanoring using 3D finite-difference time-domain method [27] (Lumerical Solutions software package, [28]). In this calculations the Ag nanoring with the abovementioned geometric parameters (see Fig. 3a) is placed on a glass substrate with a refractive index nsub = 1.45. Nanoring is irradiated from the top by the Gaussian source with a central wavelength λ = 560 nm and a full width at half maximum Δλ = 300 nm at normal incidence. To simulate the dispersion relation of silver in wide spectral range the multi-coefficient model [28] fitting the experimental data from [29] is used. As seen, in the visible spectral region good agreement is found between calculated and experimental results (Fig. 3d). The obvious difference of calculated and measured scattering spectra in the near-IR region can be explained by the excitation of lower-order plasmon modes in the case of numerical calculations [2], as well as by the absence of appropriate components in the white light source spectrum in the experiment. The absence of a sharp resonance in the visible scattering spectrum of the nanoring can be explained by the excitation of a large number of high-order plasmon modes, which resonant wavelength can be estimated using simple relation $\lambda_{sp}=2\pi R_{ring}/N$ (N is an integer number) [1]. Overlapping of their spectral curves provides the broad scattering maximum. This feature allows one to obtain enhanced electromagnetic fields near its walls. This assumption is confirmed by supporting numerical calculations of the normalized electromagnetic field amplitude $(E/E_0)^2$ ($E_0$ - the source amplitude) near the nanoring performed under irradiation of the structure by a linearly polarized Gaussian beam with the wavelengths 475 and 532 nm (see insets in Fig. 3d).

## 4. Photoluminescence enhancement on the silver nanoring

It was shown above, that the excitation of surface plasmon modes in the visible spectral range concentrates the electromagnetic energy near the nanoring walls while the constructive interference of the scattered radiation provides the focal spot inside the nanoring. In this paper, we demonstrate that such "hot spots" localized near the Ag nanoring can be used to enhance the photoluminescence (PL) signal of one of the most common organic dyes - Rhodamine 6G (R6G). For these experiments we used alcoholic solution of the R6G, which was deposited on the OF endface by means of a microinjector. Nanoring was irradiated by a CW semiconductor laser source (Milles Griot, central wavelength λ=532 nm), which was inputted into the OF section through the opposite endface (see details in the inset 2 in Fig. 1a). PL images were recorded by the high-resolution optical microscope (Hirox KH7700, optical magnification up to 7000x) equipped with a high-NA objective. The laser radiation was completely blocked by means of the long wavelength pass filter (with a band edge ~545 nm) placed in front of the microscope CCD-camera and provided the pump wavelength absorption of about $10^6$.

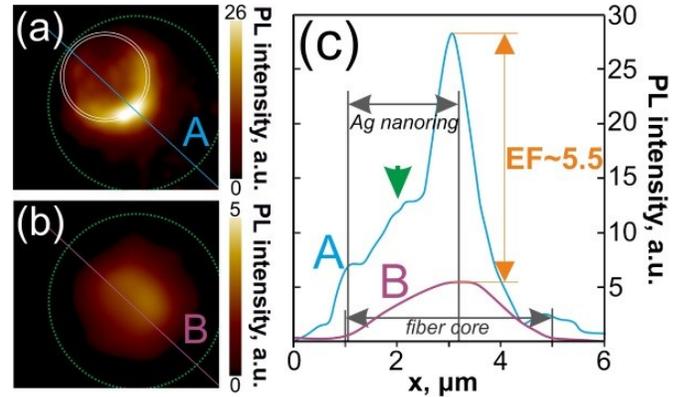

Fig. 4. Photoluminescence enhancement on the silver nanoring. Photoluminescence images of the Rhodamine 6G deposited on the endface of the optical fiber containing Ag nanoring (a) and in the case of a bare fiber core (b). Green circle marks the 4-µm-wide fiber core, while the white circles indicate the position of the Ag nanoring shifted from the fiber core center to ~1 µm. (c) Photoluminescence intensity profiles measured along directions marked with "A" and "B" letters in Figs. 4a and 4b, respectively.

Figure 4a shows the PL image of the R6G layer deposited onto the OF endface containing the Ag nanoring in the area of its optical core. As seen, high PL intensity is observed near the nanoring surface, which apparently indicates the increased electric-field intensity of the incident radiation ("hot spots") due to the plasmon mode excitation. inhomogeneity of the PL signal in our experiment is clearly defined by the asymmetric nanoring position relative to the optical core center (see Fig. 4a) and uneven irradiation of the nanoring, in turn. It also should be noted that a slight PL enhancement is observed at the nanoring center (marked with the green arrow in Fig. 4c). This enhancement is obviously associated with the scattering focal spot, which shifts to the nanoring center when laser beam irradiates the nanoring at normal incidence [5,22]. For comparison, the PL image was obtained in the case of the R6G excitation on smooth OF endface, which does not contain any irregularities in the optical core area (Fig. 4b). In this case, the PL signal is significantly weaker and uniformly distributed in the core area in accordance with the exciting radiation distribution, which indicates a key role of the nanoring in the PL intensity enhancement. By comparing the results, we have found 5.5-fold PL intensity enhancement near the silver nanoring (Fig. 4c). It should be noted that this value is slightly smaller than the enhancement factor (~ 7.5 times, [1]) achieved by using the chemically synthesized single-crystalline silver ring of almost ideal geometrical shape, which indicates high applicability of the laser-based technique for antenna-like nanoring fabrication.

## 5. Conclusions

In conclusion, we have demonstrated a simple and high-performance laser-assisted technique for silver nanoring fabrication, which includes the ablation of the Ag film by focused nanosecond pulses and subsequent reactive ion polishing. The nanoring diameter and thickness can be controlled by optimizing both the pulse energy and the metal film thickness at laser ablation step, while the subsequent reactive ion polishing provides the ability to fabricate the nanoring with desirable height. Scattering patterns of s-polarized collimated laser beam obliquely illuminating the nanoring demonstrate the focal spot inside the nanoring shifted from its center at a distance of ~ $0.57 R_{ring}$. Five-fold enhancement of the photoluminescence signal from the Rhodamine 6G organic dye near the Ag nanoring was demonstrated. This enhancement was attributed to the increase of the electromagnetic field amplitude near the nanoring surface arising from excitation of the multipole plasmon modes traveling along the nanoring. This assumption was confirmed by dark-field back-scattering spectrum of the nanoring measured under white-light illumination, as well as by supporting finite-difference time-domain simulations.

## 6. Acknowledgements

Authors are grateful for partial support to the Russian Foundation for Basic Research (Projects nos. 15-32-50996, 14-02-00205-a, 15-02-50026-a) and the project of the RAS Presidium Program. The project was also financially supported by the Russia Federation Ministry of Science and Education, Contract № 02.G25.31.0116 of 14.08.2014 between Open Joint Stock Company "Ship Repair Center "Dalzavod" and RF Ministry of Science and Education. A.A. Kuchmizhak is acknowledging for partial support from RF Ministry of Science and Education (Contract No. MK-3056.2015.2) through the Grant of RF President.